\documentclass[11pt,a4paper,twoside]{article}

\usepackage{a4}
\usepackage{amsfonts}
\usepackage{amssymb}
\usepackage{epsfig}
\usepackage{psfig}

\newcommand{\beqn}{\begin{eqnarray}}
\newcommand{\eeqn}{\end{eqnarray}}
\newcommand{\be}{\begin{equation}}
\newcommand{\ee}{\end{equation}}
\newcommand{\non}{\nonumber \\}
\newcommand{\tr}{\mbox{tr}} 
\newcommand{\Tr}{\mbox{Tr}}

\newcommand{\al}{\alpha^\prime}

\newcommand{\sz}{\scriptsize}

\begin{document} 

\title{
\begin{flushright} \vspace{-2cm} 
{\small HUB-EP-99/63 \\ \vspace{-0.15cm} 
hep-th/9912204} \end{flushright}
\vspace{3cm}
Supersymmetric 4D Orientifolds of Type IIA\\
with D6-branes at Angles
}
\author{} 
\date{}

\maketitle

\begin{center}
Ralph Blumenhagen,\footnote{e-mail: blumenha@physik.hu-berlin.de} 
Lars G\"orlich\footnote{e-mail: goerlich@physik.hu-berlin.de} and  
Boris K\"ors\footnote{e-mail: koers@physik.hu-berlin.de} \\
\vspace{0.5cm}
Humboldt Universit\"at zu Berlin \\
{\small{Institut f\"ur Physik, Invalidenstr. 110, 10115 Berlin, Germany}} \\
\vspace{2cm}
\end{center}

\begin{abstract}
\noindent
We study a certain class of four-dimensional ${\cal N}=1$ supersymmetric
orientifolds for which the world-sheet parity transformation
is combined with a complex conjugation in the compact
directions. We investigate  in detail the orientifolds of the $\mathbb{Z}_3$, 
$\mathbb{Z}_4$, $\mathbb{Z}_6$ and $\mathbb{Z}_6^\prime$  toroidal orbifolds
finding solutions to the tadpole cancellation conditions
for all models. Generically, all the massless spectra turn out to
be non-chiral.

\end{abstract}

\thispagestyle{empty}
\clearpage

\tableofcontents

\section{Introduction}

Open string models have largely extended our view on consistent 
string theory vacua. The foremost promising class of heterotic 
Calabi-Yau compactifications has become only a small subset
of string vacua featuring ${\cal N}=1$ space time supersymmetry
in four dimensions and allowing chiral spectra in phenomenological
interesting gauge groups. Non-perturbative heterotic vacua containing
solitonic five-branes in the background are better described by
their dual type I models, as the heterotic five-brane is mapped
to the D5-brane in type I. 
Generally, type I models  contain D-branes in the background supporting
the gauge sector of the low energy theory, whereas gravity propagates 
in the ten-dimensional bulk. Therefore, such models became
quite attractive in recent attempts to establish a unification
scenario with the string scale as low as 1TeV \cite{rtev}. 

Type I models can be described  as orientifolds of the type IIB 
superstring by the world-sheet parity transformation $\Omega$.
Compactification of type I string theory on various toroidal
orbifolds in six and four space dimensions have been studied
in a couple of papers \cite{rsagbi}-\cite{rkakunp}. In particular, 
for four dimensional
compactifications it was found that it is not always possible to 
satisfy all tadpole cancellation conditions, showing a perturbative
inconsistency of some of the models \cite{rkakunp}.   

In \cite{hepth9908130} we presented a new kind of six-dimensional
orientifolds, for which the world-sheet parity transformation is combined
with a complex conjugation ${\cal R}$ of the internal coordinates.
In order to cancel the tadpoles appearing in the Klein bottle amplitude
it was necessary to introduce D7-branes intersecting at non-trivial angles
into the background. Thus, for the $\mathbb{Z}_3$, $\mathbb{Z}_4$ and
$\mathbb{Z}_6$ toroidal orbifolds we found consistent $\Omega{\cal R}$
orientifolds in six dimensions leading indeed to anomaly free massless
spectra. Generally, the rank of the gauge groups turned out to be reduced
by powers of two as compared to the non-compact theory 
and extra multiplicities appeared in what we called
twisted open string sectors. Note, that similar models were also discussed in
\cite{bimopra}, however the precise
relation between our models and the models in \cite{bimopra} is not
clear. 

In \cite{AngBlu} T-duality was used to gain a better
understanding of the rank reduction for $\Omega\cal{R}$ orientifolds. 
In contrast to standard $\Omega$ orientifolds, it turned out that for
$\Omega\cal{R}$ orientifolds the background antisymmetric two-form field
is a continuous modulus, whereas the off-diagonal part of the internal
metric is frozen to discrete values. Turning on these discrete
parameters leads to the reduction of the  rank of the gauge group
by powers of two. 
In this paper we will present a geometric understanding of the extra 
multiplicities in twisted open string sectors, which we introduced by hand
in \cite{hepth9908130}. From the loop channel
point of view these factors are simply intersection numbers of the
D-branes in question. 

In the following we will study supersymmetric 
$\Omega{\cal R}$ orientifolds of type IIA in four space-time dimensions. 
We focus our attention on toroidal orbifolds preserving ${\cal N}=1$ 
supersymmetry and study the $\mathbb{Z}_3$, $\mathbb{Z}_4$,
$\mathbb{Z}_6$ and  $\mathbb{Z}_6^\prime$ models in detail. 
Analogous to the six-dimensional case, 
the Klein bottle tadpoles have to  be cancelled by introducing D6-branes
at non-trivial angles into the background. Depending on the
defining lattices for the  three $T^2$ tori, we find that there 
exists more than one consistent choice for each $\mathbb{Z}_N$ orbifold.
In particular, in contrast to ordinary orientifolds, we find solutions
to the tadpole conditions for each $\mathbb{Z}_N$ orbifold. 
Moreover, for the rank of the resulting gauge groups we always obtain 
exactly the reduction expected from the general arguments in \cite{AngBlu}.
At first sight surprisingly, we only get non-chiral massless spectra,
where in the two cases, $\mathbb{Z}_4$ and  $\mathbb{Z}_6$, 
accidentally the massless open string spectra fit 
into ${\cal N}=2$ supermultiplets. 

This paper is organized as follows. In section 2 we describe the general 
features of $\Omega{\cal R}$ orientifolds in four dimensions. In section
3 we discuss the $\mathbb{Z}_4$ example in some detail, as it already exhibits
all the technical steps one has to go through when computing such models.  
The sections 4 and 5 summarize  the results of our computation for
the $\mathbb{Z}_3$ and the two $\mathbb{Z}_6$ examples, respectively. 
In an appendix we have listed the Klein bottle, annulus and M\"obius strip
amplitudes for all models discussed in this paper. In the  conclusions
we finish with a brief discussion of non-supersymmetric generalizations
of the $\Omega{\cal R}$ orientifolds.

\section{$\Omega{\cal R}$ orientifolds}

We consider orientifolds which are obtained by combining the 
ordinary world sheet parity transformation $\Omega$ with a conjugation 
${\cal R}$ 
of the three complex coordinates of the six-dimensional torus $T^6$.
Since ${\cal R}$ can also be considered as a reflection of three of the six
compact coordinates, $\Omega{\cal R}$ is rather a symmetry of the type IIA 
than of the
type IIB superstring. This operation is accompanied by one of the well 
known cyclic orbifold groups preserving ${\cal N}=1$ supersymmetry in 
four space-time dimensions. Thus, the entire orientifold group is given 
by $\mathbb{Z}_N \cup \Omega{\cal R} \mathbb{Z}_N$. 
Note, that these models 
are not related by T-duality to standard orientifolds
$\mathbb{Z}_N \cup \Omega \mathbb{Z}_N$. Instead they are dual to standard 
asymmetric orientifolds $\hat\mathbb{Z}_N \cup \Omega 
\hat\mathbb{Z}_N$ which we think deserve to be studied in their
own right. To begin with we give a more explicit definition of all the 
elements required.

\subsection{Proper definition}

The worldsheet parity transformation $\Omega$ is combined with a 
particular reflection ${\cal R}$, 
which in terms of the complex coordinates
\beqn
X_i \equiv x_{10-2i}+i x_{11-2i},\quad i=1,2,3,
\eeqn
of a six-dimensional torus can be written as the conjugation
\beqn
{\cal R}: X_i \mapsto \bar{X}_i.  
\eeqn
In this basis the generator $\Theta$ of the orbifold group
$\mathbb{Z}_N=\{ 1,\Theta, ...,\Theta^{N-1} \}$ 
acts diagonally 
\beqn
\Theta : X_i \mapsto \exp \left( 2\pi i v_i \right) X_i
\eeqn
and with opposite phases on the conjugate variables. 
In the same way the complexified fermionic coordinates diagonalize $\Theta$. 
Both for the RR and the twisted NSNS sector the operation of ${\cal R}$ and
$\Theta$ on degenerate groundstates is given by
\beqn \label{Rongroundst}
{\cal R} : \vert s_0, s_1, s_2, s_3 \rangle & \mapsto & 
\vert s_0, -s_1, -s_2, -s_3 \rangle , \non
\Theta :  \vert s_0, s_1, s_2, s_3 \rangle & \mapsto & 
 \exp \left( 2\pi i \vec v \cdot \vec s \right) 
\vert s_0, s_1, s_2, s_3 \rangle .
\eeqn
If the groundstate is not a spinor of the entire light-cone gauge $SO(8)$ 
little group, but only of a $SO(2k)$ 
subgroup, one can formally set the respective $s_i$ in (\ref{Rongroundst}) to 
zero to obtain the correct transformation. 
There is a subtlety concerning the GSO projection in the twisted sectors. 
Taking into account that  the $\Omega{\cal R}$ projection is related to the
$\Omega$ projection by T-duality in the three directions in which
${\cal R}$ acts non-trivially, one is led to the following left moving and
right moving world sheet fermion number operators 
\beqn \label{fermionn}
\left( -1\right)^{F_L} \vert s_0, s_1, s_2, s_3 \rangle
&=& \exp \left( 2\pi i(s_0 -s_1 -s_2- s_3) \right) 
\vert s_0, s_1, s_2, s_3 \rangle, \non
\left( -1\right)^{F_R} \vert s_0, s_1, s_2, s_3 \rangle
&=& \exp \left( 2\pi i(s_0+ s_1 +s_2+ s_3) \right) 
\vert s_0, s_1, s_2, s_3 \rangle.
\eeqn
With this choice of world-sheet fermion number operators 
the GSO projection is on states satisfying 
$(-1)^{F_L}=(-1)^{F_R}=-1$.
In the untwisted sector of the orientifolds 
this is apparently  equivalent to the usual type IIA GSO projection
and in the twisted sectors it guarantees that $\Omega{\cal R}$ is really
a symmetry of the type IIA orbifold theory. In the various open string sectors
the GSO projection is always determined by supersymmetry.\\

Orbifolds that allow ${\cal N} =1$ supersymmetry in four dimensions have been 
classified in \cite{DHVW1985,DHVW1986} 
and we display their action on the complex basis in terms of the 
$v_i$ in Table \ref{tablezn}. \\

\begin{table}[h]
\caption{$\mathbb{Z}_N$ groups that preserve ${\cal N}=1$ in $d=4$}
\label{tablezn}
\begin{center} 
\begin{tabular}{|l|l|l|}
\hline
$\mathbb{Z}_3 : v= (1,1,-2)/3$ & $\mathbb{Z}^\prime_6 : v= (1,2,-3)/6$ &
$\mathbb{Z}_8^\prime : v= (1,2,-3)/8$ \\
$\mathbb{Z}_4 : v= (1,1,-2)/4$ & $\mathbb{Z}_7 : v= (1,2,-3)/7$ &
$\mathbb{Z}_{12} : v= (1,4,-5)/12$ \\
$\mathbb{Z}_6 : v= (1,1,-2)/6$ & $\mathbb{Z}_8 : v= (1,3,-4)/8$ &
$\mathbb{Z}_{12}^\prime : v= (1,5,-6)/12$ \\
\hline
\end{tabular}
\end{center}
\end{table}

\noindent
In this paper we will restrict ourselves to explicit computations for
the orbifolds 
$\mathbb{Z}_3$, $\mathbb{Z}_4$, $\mathbb{Z}_6$ and 
$\mathbb{Z}_6^\prime$.
It has been established for ordinary four dimensional type I  
vacua that there exists no solution to the tadpole cancellation
conditions for the cases $\mathbb{Z}_4$, $\mathbb{Z}_8$, $\mathbb{Z}_8^\prime$ 
and $\mathbb{Z}_{12}$ \cite{riban} \footnote{See also
\cite{Susybreak} for a recent discussion of similar models leading
to supersymmetry breaking on the D-branes.}.
In contrast to that, we obtain perturbatively consistent $\Omega{\cal R}$ 
orientifolds for any $\mathbb{Z}_N$ model studied in this paper. \\

For the compact $T^6$ one needs to use 
a lattice that allows a crystallographic action of the above cyclic groups.
Up to overall scales in each $T^2$ factor, for the $\mathbb{Z}_3$ and 
$\mathbb{Z}_6$ case we choose the root lattice of $SU(3)^3$ and for
the  $\mathbb{Z}_4$ case the root lattice of $SU(2)^6$.
Except for the $\mathbb{Z}_3$ example these are not 
of the type explored in \cite{MOP1987}, which allow to take the coxeter 
element of  the Lie algebra as the generator of the orbifold group. 
As the particular choice of the lattice enters at various important points 
into the calculation of the tadpole cancellation 
conditions and the spectrum, different lattices may lead to even more 
inequivalent models. 
With our choice we obtain twelve inequivalent models from the four orbifolds by 
the freedom of choice in the relative orientations of the lattices with 
respect to the reflection ${\cal R}$.\\ 
 
In the following we describe in some detail the general features that arise 
in computing the three contributions to the massless tadpoles arising in 
the Klein bottle, annulus  and M\"obius strip amplitudes.
The truly technical details of the explicit expressions can be found in  
the appendix. Part of the following discussion already appeared 
in \cite{hepth9908130}, 
but in a couple of important questions substantial progress 
could be achieved.
For instance, we clarify  the appearance of extra weight factors for twisted 
open string sectors in the annulus and M\"obius strip amplitudes.
In \cite{hepth9908130} these factors were detected by employing the
equivalence between the tree channel and the loop channel description
of the annulus.
Now, we can give a striking geometric interpretation of these factors
as intersection numbers of the D-branes supporting the twisted open
string sector.

\subsection{Closed strings}

In the closed string sector the states with excitations of 
left-moving oscillators combine 
with their complex conjugate images on the right-moving side into 
$\Omega{\cal R}$ invariant states 
\beqn
\label{ominv}
\left( \Omega{\cal R}\right) \left(\alpha_{n+kv_i} 
\tilde{\overline{\alpha}}_{n+kv_i}
\right)\left( \Omega{\cal R}\right)^{-1}  & = &
\left(\alpha_{n+kv_i} \tilde{\overline{\alpha}}_{n+kv_i}\right) , \non 
\left( \Omega{\cal R}\right) \left(\overline{\alpha}_{n-kv_i} 
\tilde{\alpha}_{n-kv_i}
\right)\left( \Omega{\cal R}\right)^{-1}  & = &
\left(\overline{\alpha}_{n-kv_i} \tilde{\alpha}_{n-kv_i}\right) 
\eeqn
with $\alpha$ representing any kind of bosonic or fermionic ladder operator of the 
$k$-twisted sector and $n\in \mathbb{Z}$ or $n\in \mathbb{Z}+1/2$.
As opposed  to standard orientifolds the $\Omega{\cal R}$ invariant 
states in (\ref{ominv}) are also invariant under the $\mathbb{Z}_N$ action
\beqn
\label{znact}
\Theta \left(\alpha_{n+kv_i} \tilde{\overline{\alpha}}_{n+kv_i}\right) 
\Theta^{-1}  & = & 
\left(\alpha_{n+kv_i} \tilde{\overline{\alpha}}_{n+kv_i}\right), \non
\Theta \left(\overline{\alpha}_{n-kv_i} \tilde{\alpha}_{n-kv_i}\right) 
\Theta^{-1}  & = & 
\left(\overline{\alpha}_{n-kv_i} \tilde{\alpha}_{n-kv_i}\right).
\eeqn
The important point to notice is that 
$\Omega{\cal R}$ does not exchange a $\Theta^k$ twisted sector with a
$\Theta^{N-k}$ twisted sector, so that all twisted sectors lead to
a non-vanishing contribution to the loop channel Klein bottle amplitude
\beqn
\label{kleinloop}
{\cal K} = 4c \int_0^\infty{\frac{dt}{t^3}\ \Tr_{\mbox{\sz U+T}} 
\left( \frac{\Omega{\cal R}}{2} 
\frac{\left( 1+\Theta+ \cdots +\Theta^{N-1} \right)}{N} 
\frac{\left( 1+(-1)^F \right) }{2} e^{-2\pi t \left( L_0 + \bar{L}_0 \right) }
\right) } ,
\eeqn
with $c \equiv V_4/ \left( 8\pi^2 \al \right)^2$ 
and the momentum integration in the non-compact space-time already performed.
Remember, that for conventional orientifolds only the untwisted and, if present, the 
$\mathbb{Z}_2$ twisted sectors contribute to this amplitude. 
Here, the relation 
\beqn
\Omega{\cal R}\Theta^k = \Theta^{N-k}\Omega{\cal R}
\eeqn
implies that in the tree channel only untwisted closed string states  
propagate between the two cross-caps on both sides of the tube.
Thus, world sheet consistency requires that transforming the loop
channel amplitude (\ref{kleinloop}) into tree channel must lead
to an amplitude in which only $\mathbb{Z}_N$ invariant states 
from the untwisted sector
contribute. In doing the computation one realizes that this world-sheet
consistency condition is not always automatically satisfied.
Therefore, the completion of  the $\mathbb{Z}_N$ projector in the tree channel
will serve as  the guiding principle in constructing consistent
models. \\
 
Since the action of $\Theta$ in the basis of $\Omega{\cal R}$ invariant 
states  is always trivial on the oscillator part,  in each $\mathbb{Z}_N$ 
twisted sector the complete Klein
bottle amplitude factorizes into a trace over the oscillators times 
a trace over the lattice.
To compute the lattice part, one needs to determine the winding (W) and the
Kaluza-Klein (KK) modes which are invariant 
under the operator $\Omega{\cal R}\Theta^k$ appearing in the trace. 
To begin with, let us consider a single two-dimensional torus, on which
$\Theta$ acts as a rotation by an angle $\phi=2\pi /N$.
In fact the entire lattice partition function of the six-dimensional torus 
factorizes into a product of three two-dimensional tori with an action of
$\Theta$ as a $2\pi v_i/N$ rotation on each individual factor.  
Let us define the lattice type {\bf A} to be the orientation of the 
$\mathbb{Z}_N$ lattice such that the reflection ${\cal R}$ acts 
orthogonally to one of the two basis vectors that 
span the lattice. In this case the relation
\beqn
\left( \Omega{\cal R} \Theta^k \right) \Theta^2 = \Theta^{-1} 
\left( \Omega{\cal R} \Theta^k \right) \Theta
\eeqn  
implies that the partition function 
of the lattice with $\Omega{\cal R}\Theta^k$ insertion does only depend on $k$ 
being even or odd.
Rotating the lattice by an angle of $\pi /N$  for even and $\pi /(2N)$ for odd $N$  
leads to the lattice type
{\bf B} and constitutes  the only non-trivial rotation maintaining  
a crystallographic action of ${\cal R}$. Due to the relation
\beqn
\Theta^{-1/2} \left( \Omega{\cal R} 
\Theta^k \right) \Theta^{1/2} = \left( \Omega{\cal R} \Theta^k \right) \Theta 
\eeqn 
the two cases, even and odd, get exchanged. 
Note, that in the special case of the $\mathbb{Z}_3$ orbifold
the rotation $\Theta^{1/2}$ is a symmetry of the lattice
implying that the Klein bottle contributions are  entirely independent of $k$.

It appears to be the correct choice for world-sheet consistency, that
whenever the orbifold model has two complex directions with 
$Nv_i$ odd and only one with $Nv_i$ even, ${\cal R}$ acts differently 
on the two 
odd directions and arbitrarily on the even direction.
Said differently, we choose the lattices {\bf AB} or {\bf BA} for the
two odd directions and in the even direction we are free to choose
either {\bf A} or {\bf B}.  
In this way one produces a partition function 
which in the untwisted sector loop channel yields  identical contributions 
for all $\Theta^k$ insertions in the trace. Accidentally, 
this remains true for the twisted sectors with the exception of  the
$\mathbb{Z}_6^\prime$ orientifold.  
In the $\mathbb{Z}_6^\prime$ case, the two choices {\bf AB} and {\bf BA}
for the two odd tori lead to different models.
Summarizing, for the $\mathbb{Z}_3$ orbifold we have the four inequivalent
choices $\{{\bf AAA},{\bf AAB},{\bf ABB},{\bf BBB}\}$, for the
$\mathbb{Z}_4$ and the $\mathbb{Z}_6$ the two choices $\{{\bf ABA},{\bf ABB}\}$
and finally for the $\mathbb{Z}_6^\prime$ the four choices
$\{{\bf AAB},{\bf ABB},{\bf BAA},{\bf BBA}\}$.\\

We have also inspected some of the cases without any relative angle 
between the lattices of the two-dimensional tori, but found that these 
never lead to the complete  $\mathbb{Z}_N$ projector in the 
tree channel. Thus, the world-sheet consistency condition is violated 
and in the spirit of \cite{rkakunp} it is tempting to speculate 
that this might hint to the existence of non-perturbative states which  
remove this discrepancy. 
To prove these speculation it is of primary importance to find
a heterotic or F-theory dual description of the $\Omega{\cal R}$ 
orientifolds.\\

In order to compute the loop channel Klein bottle amplitude in the $\Theta^l$ 
twisted sector, we have to know which of the $\Theta^l$
fixed points are invariant under the action of $\Omega{\cal R} \Theta^k$.
These multiplicities arise from 
the center of mass coordinate of the string. It is required 
to be invariant under 
the twisting element as well as under the insertion in the trace. 
With the exception of the $\mathbb{Z}_6^\prime$ orientifold
the number of invariant fixed points does not depend on 
$k$, though different individual points are  invariant. 
In the $\mathbb{Z}_6^\prime$ example the $\Theta^2$ and $\Theta^4$ twisted
sectors split into a sum of two terms containing the
$\Omega{\cal R} \Theta^{2k}$ and $\Omega{\cal R} \Theta^{2k+1}$ insertions
in the trace, respectively. 
Note, that the fixed point multiplicities non-trivially 
conspire with the relative factors 
arising  from the modular transformation of the 
lattice contributions in order to complete the tree channel 
projector. \\

Regarding these subtleties the computation of the various amplitudes 
is a straightforward though still tedious task. All the results for the 
different traces involved can be found in the appendix and the
$\mathbb{Z}_4$ example will be discussed in greater detail in the following 
chapter. \\

Finally, we need to  compute the massless spectra. The first step is to 
add up the ground state energies in the twisted sectors 
using the general  formula 
\beqn
E_0 = \sum_{i=0}^3{\left( \frac{1}{24} - \frac{1}{8} 
\left( 2kv_i -1\right)^2 \right)} ,
\eeqn
which is presented for a complex boson of the $\Theta^k$-twisted sector.
For the fermions the sign changes to $-E_0$ and in the NSNS sector one
needs to 
replace $v_i$ by $v_i+1/2$. In the untwisted sector $\Theta$ invariant
left- and right-moving massless states have to be 
symmetrized and antisymmetrized under $\Omega{\cal R}$ in the NSNS and the 
RR sector respectively. 
This always contributes the graviton and dilaton from the NSNS sector and some 
model dependent number of additional neutral chiral multiplets.
For the twisted sectors one carefully needs to inspect the transformation 
properties of  the fixed points. For instance those which are invariant 
under ${\cal R}$ and $\Theta$ need to be properly symmetrized and 
antisymmetrized 
as in the untwisted sector.  It also happens that the fixed points are
only invariant under a combination of ${\cal R}$ and $\Theta$ or even 
under neither of the two transformations.
In the latter case no symmetrization at all has to take place. 
Following this procedure, the total number of chiral (C) and vector (V) 
multiplets\footnote{We 
would like to acknowledge that \cite{hepth9912218} has drawn our attention to 
the necessity of further distinguishing states in the twisted RR sectors, that 
transform as scalars respectively as vectors.} 
is exactly the sum of the Hodge numbers of the blown up 
toroidal orbifold
\beqn
{\mbox{Number of neutral multiplets C+V}}\  = h^{1,1}+h^{2,1} .
\eeqn
In particular, for inequivalent orientifold models with 
identical orbifold groups 
we find the same net number of multiplets in the closed string spectrum, 
while the individual states are different. 
In Table \ref{tableclosed} we display the closed string spectra 
for the orientifolds discussed in this paper. \\

\begin{table}[h]
\caption{Closed string spectra}
\label{tableclosed}
\begin{center} 
\begin{tabular}{|l|l|c|c|c|c|}
\hline
Orbifold & Model & untwisted & 
        $\Theta+\Theta^{-1}$  & $\Theta^2 \left(+\Theta^{-2}\right)$  & 
        $\Theta^3$  \\ 
group & & & twisted & twisted & twisted\\
\hline 
$\mathbb{Z}_3$ & {\bf AAA} & 9{\mbox C} & 14{\mbox C}+13{\mbox V} & absent & absent \\
$\mathbb{Z}_3$ & {\bf AAB} & 9{\mbox C} & 15{\mbox C}+12{\mbox V} & absent & absent \\
$\mathbb{Z}_3$ & {\bf ABB} & 9{\mbox C} & 18{\mbox C}+9{\mbox V} & absent & absent \\
$\mathbb{Z}_3$ & {\bf BBB} & 9{\mbox C} & 27{\mbox C} & absent & absent \\
$\mathbb{Z}_4$ & {\bf ABA} & 6{\mbox C} & 16{\mbox C} & 15{\mbox C}+1{\mbox V} & absent \\
$\mathbb{Z}_4$ & {\bf ABB} & 6{\mbox C} 
& 12{\mbox C}+4{\mbox V} & 15{\mbox C}+1{\mbox V} & absent \\
$\mathbb{Z}_6$ & {\bf ABA} & 5{\mbox C} & 2{\mbox C}+1{\mbox V} 
& 9{\mbox C}+6{\mbox V} & 10{\mbox C}+1{\mbox V} \\
$\mathbb{Z}_6$ & {\bf ABB} & 5{\mbox C} & 3{\mbox C} 
& 12{\mbox C}+3{\mbox V} & 10{\mbox C}+1{\mbox V} \\
$\mathbb{Z}_6^\prime$ & {\bf AAB} & 4{\mbox C} & 7{\mbox C}+5{\mbox V} 
& 14{\mbox C}+4{\mbox V} & 10{\mbox C}+2{\mbox V} \\
$\mathbb{Z}_6^\prime$ & {\bf ABB} & 4{\mbox C} & 9{\mbox C}+3{\mbox V} 
& 18{\mbox C} & 10{\mbox C}+2{\mbox V} \\
\hline 
\end{tabular}
\end{center}
\end{table}

\noindent
In the untwisted sector there are always the graviton and the dilaton multiplet 
in addition to the chiral multiplets as given in the table.

\subsection{Open strings}

Analogously to the six-dimensional models studied in \cite{hepth9908130},
in order to cancel the tadpoles from the Klein bottle 
we have to introduce D6-branes intersecting each other at non-trivial angles. 
They always 
stretch in one direction on each two-dimensional torus and need to intersect 
in a $\Theta$ and ${\cal R}$ invariant fashion. 
Matching the modings of the fields in the closed string
twisted sectors, open strings stretching between D-branes 
at  relative angle $\pi v_i$ carry fields with modings shifted by $v_i$.
Thus, we are led to consider arrays of $N$ kinds of D-branes at 
this relative angle.  
On each torus $T^2$  one of them is located in the fixed plane of the 
reflection ${\cal R}$ and the remaining ones are obtained by successively  
applying the rotation $\Theta^{1/2}$.     
Note that some of them will coincide in some complex directions eventually. \\

In the open string sector one has to compute the annulus 
\beqn
{\cal A}= c\int_0^\infty{\frac{dt}{t^3}\ \Tr_{\mbox{\sz open}} 
\left( 
        \frac{1}{2} \frac{1+\Theta +\cdots +\Theta^{N-1}}{N} 
        \frac{ 1+(-1)^F}{2} e^{-2\pi tL_0}
\right) }
\eeqn 
and the M\"obius strip amplitude 
\beqn
{\cal M}= c\int_0^\infty{\frac{dt}{t^3}\ \Tr_{\mbox{\sz open}} 
\left( 
        \frac{\Omega{\cal R}}{2} \frac{1+\Theta +\cdots+\Theta^{N-1}}{N} 
        \frac{ 1+(-1)^F}{2} e^{-2\pi tL_0} 
\right) .}
\eeqn
These amplitudes only receive a non-vanishing 
contribution from open strings in the $\left( 6_i,6_{i+n} \right)$ sector, 
if the  action of the operator in the trace leaves the two D6-branes and
the orientation of the open string invariant. 
If the moding of the fields in the $\left( 6_i,6_{i+n} \right)$ 
open string sector
is similar to the moding of the fields in the $\Theta^k$ 
twisted closed string
sector, we also use the term ``$\Theta^k$ twisted'' for the open string sector.
For even $N$ the operator $\Theta^{N/ 2}$ leaves all D6$_i$-branes 
invariant and has a non-trivial action on the Chan-Paton factors as usually
described by a matrix $\gamma_{N/2}^{(i)}$. Correspondingly, 
the annulus amplitude gives rise to an additional $\mathbb{Z}_2$ 
twisted sector tadpole. \\

Some care has to be taken when computing the contributions of momenta and 
winding states, 
which are present for $\left( 6_i,6_{i+n} \right)$ strings whenever the two
branes coincide in some complex direction and the operator in the trace 
acts trivially there. One then needs to consider the orientation of the 
brane with respect to the lattice and determine the allowed winding and 
momentum states. In the M\"obius strip amplitude these KK and W modes
also have to be invariant under $\Omega{\cal R}$. Generally this leads to a  
doubling of winding states as compared to the annulus,
with the exception  of the {\bf A} type $SU(2)^2$ lattice. The 
present choice of lattices and branes guarantees that the contributions are 
independent of the operator in the trace for a given sector, except again
for the   $\left( 6_i,6_{i+2} \right)$ strings in the $\mathbb{Z}_6^\prime$. \\

A very important point in the computation are the extra multiplicities of some 
twisted open string 
sectors introduced in \cite{hepth9908130} by hand in order to 
satisfy the world-sheet consistency condition.
These extra multiplicities were determined by the complete projector in 
the tree channel and  nicely led to tadpole cancellation and anomaly 
free massless spectra. Unfortunately, in \cite{hepth9908130} we could
not present an understanding of these factors from the loop channel
point of view, which we would like to do now. 
The source for the extra multiplicities 
is again the center of mass coordinate of the open string, 
which is required to be an 
intersection point of the two D-branes. The number of such intersection points 
of the respective D-branes in question perfectly reproduces 
these extra factors and thus gives a striking geometrical interpretation. 
For computing the annulus and M\"obius strip amplitude, these intersection
points  need to be invariant under the operator in the trace. 
It turns out that the number of invariant intersection points of 
$\left( 6_i,6_{i+n} \right)$ branes is independent of $i$ with the exception 
of $n=2$ in the $\mathbb{Z}_6^\prime$ orientifold. 
In \cite{AngBlu} these extra multiplicities have been related via T-duality to 
extra  factors for twisted sectors in ordinary $\Omega$ orientifolds 
with background 
$B^{\mu \nu}$ field.  
So far we have presented all the novel ingredients needed to compute the 
two open string amplitudes and the details are left to the appendix and 
to the discussion of the $\mathbb{Z}_4$ example. \\ 

Let us now discuss the main steps for computing the massless open string 
spectrum. One has to be very careful again with the contributions 
of the different intersection points. First of all we notice,
that one always gets a tadpole cancellation condition of the form
\beqn
({\rm M}-2^\kappa)^2=0
\eeqn 
fixing the number ${\rm M}$ of D6-branes of each type. For even $\mathbb{Z}_N$
there is an additional  $\mathbb{Z}_2$ twisted  tadpole condition, 
which requires 
\beqn
\tr \left( \gamma_{N/2}^{(i)} \right)=0 
\eeqn 
exactly resembling the computation for the standard
six-dimensional $\mathbb{Z}_2$ orientifold discussed by  
Gimon and Polchinski (GP) in \cite{rgimpol}. 
We can therefore copy their solution for the Chan-Paton degrees of freedom,
which implies an $SO({\rm M})$ gauge group on each stack of branes, 
broken to its $U({\rm M}/2)$ subgroup, if $N$ is even. 
We simplify the analysis by looking at those open strings only, 
which begin on one of the branes located in the 
fixed plane of ${\cal R}$, as all other states are related to these by some 
action of the orientifold group. 
Again, for the $\mathbb{Z}_6^\prime$ model the situation is slightly
different and one has to consider the branes of odd and even $i$ 
separately. 
 
For the massless states in the untwisted NS sector 
the action of the ordinary $\Omega$ on D9-branes is identical to 
the action of $\Omega{\cal R}$ on D6-branes, 
as the additional signs which $\Omega$ contributes to the Dirichlet directions 
just cancel the reflection. 
For even $N$ we therefore only need to distinguish further,
whether $\Theta^{N/2}$ acts by a reflection or trivially on a given state, 
i.e. if $Nv_i/2$ is integer or not. A single  
physical state with a reflection contributes two states of a multiplet in the 
antisymmetric representation of the gauge group, 
while a state with trivial action 
contributes a single state of a multiplet in the adjoint representation. 
By inspecting Table \ref{tablezn} one can easily realize, 
that generically there is 
one complex direction, which $\Theta^{N/2}$ acts trivially on, 
while it reflects the other two. 
By this reasoning we find, that the untwisted sector always contributes a
vector and a chiral multiplet in the adjoint representation 
of the gauge group from  the trivial directions and four chiral multiplets
in the antisymmetric representation from the reflected directions.  
For odd $N$, in addition to the vector multiplet there simply appear
3 chiral multiplets in the antisymmetric representation.

Similar to the closed string case, for the twisted sectors things 
become more involved. First 
one needs to do the same distinction concerning the Chan-Paton labels as 
for the 
untwisted sector above. Moreover, one has to distinguish the 
contributions of 
the various intersection points, which may be invariant under $\Theta$ 
and ${\cal R}$ or not. Finally, one must keep track of various phase factors,
that appear via modular transformation and provide additional relative
signs. 
This needs to be taken into
account, when doing any symmetrization or antisymmetrization of 
Chan-Paton labels. 
Since the computation of the open string spectrum of the
four-dimensional $\Omega{\cal R}$ orientifolds is similar to a $\mathbb{Z}_2$
standard orientifold, we only find non-chiral spectra in mostly rather small
gauge groups. Remember, that standard four-dimensional $\mathbb{Z}_N$
orientifolds for $N>2$ generically have chiral spectra. \\

After having explained all the principles, we now come to the
detailed discussion of one example. 
We have chosen the $\mathbb{Z}_4$ as it already exhibits 
all the generic features we have went through while not being as 
complicated in technical terms as the $\mathbb{Z}_6^\prime$.

\section{The $\mathbb{Z}_4$ orientifold}

\begin{figure}[h]
\begin{center}
\makebox[10cm]{
 \epsfxsize=12cm
 \epsfysize=6cm
 \epsfbox{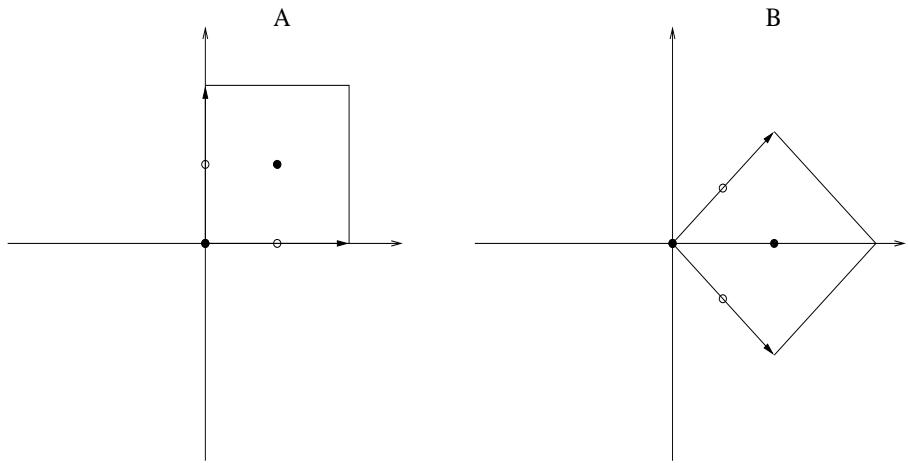}
}
\end{center}
\caption{The $\mathbb{Z}_4$ lattices}
\label{figz4lattice}
\end{figure}

The $\mathbb{Z}_4$ action is given by $v=(1,1,-2)/4$ and up to scales
for the $T^6$ we choose the root lattice of $SU(2)^6$.   
In Figure \ref{figz4lattice} we depicted the two 
possible choices of  relative orientations of the $SU(2)^2$ lattice
with respect to the fixed line of ${\cal R}$, which is identical to
one of the coordinate axes for each $T^2$.\\ 
%%%%%%%%%%%%%%%%%%%%%%%%%%%%%%%%%%%%%%%%%%%%%%%%%%%%%
%   BILD: 
%%%%%%%%%%%%%%%%%%%%%%%%%%%%%%%%%%%%%%%%%%%%%%%%%%%%%
%%%%%%%%%%%%%%%%%%%%%%%%%%%%%%%%%%%%%%%%%%%%%%%%%%%%%
%
%%%%%%%%%%%%%%%%%%%%%%%%%%%%%%%%%%%%%%%%%%%%%%%%%%%%%

The KK and W states invariant under $\Omega{\cal R} \Theta^{2k}$ are 
\beqn
\label{kkwinv}
p^{\bf A} &=& \frac{m}{R},\ l^{\bf A} = nR \non
p^{\bf B} &=& \sqrt{2} \frac{m}{R},\ l^{\bf B} = \sqrt{2} nR ,
\eeqn
for the respective orientations of the lattice. 
The states invariant under $\Omega{\cal R} \Theta^{2k+1}$ are 
obtained by exchanging ${\bf A}$ and ${\bf B}$ in (\ref{kkwinv}) for the first 
two $SU(2)^2$ tori, 
whereas they are identical to (\ref{kkwinv})  for the third  $T^2$. 
By world-sheet consistency we are forced to choose the lattices 
in the first two $T^2$ tori as {\bf AB} and in the third $T^2$ we are
free to choose it either {\bf A} or {\bf B}.
Thus, we get the two inequivalent models {\bf ABA} and {\bf ABB}.
We display all results 
first for the {\bf ABA} model and state relative factors for the {\bf ABB}
case separately. The number of fixed points in  $\Theta$  and $\Theta^3$ 
twisted sectors is 16 in either case, but for 
{\bf ABB} only 8 are invariant under $\Omega{\cal R} \Theta^k$.
In the $\Theta^2$ twisted sector one has 
also 16 fixed points, of which one half is invariant under 
$\Omega{\cal R} \Theta^k$ both for {\bf ABA} and {\bf ABB}. 
Note, that the expected relative prefactors for the terms in the complete 
projector in the tree channel amplitude are 
\beqn
\label{pref}
\Theta^{1} &:& \quad \left( -2\sin \left( \frac{\pi}{4} \right) 
\right)^2 \left( -2\sin \left( -\frac{\pi}{2} \right) \right) =4 , \non
\Theta^2 &:& \quad \left( -2\sin \left( \frac{\pi}{2} \right) 
\right)\left( -2\sin \left( -\frac{\pi}{2} \right) 
\right) = -4,\non
\Theta^{3} &:& \quad \left( -2\sin \left( \frac{3\pi}{4} \right) 
\right)^2 \left( -2\sin \left( -\frac{3\pi}{2} \right) \right) =-4.
\eeqn
as explained in the appendix. 
This will be found to be consistent for both  {\bf ABA} and {\bf ABB}.\\

After this fixed point analysis we have all the ingredients needed 
to compute the Klein bottle trace 
\beqn
{\cal K} &=& c\ \int_0^\infty{\frac{dt}{t^3}\ 
\left( 
{\cal K}^{(0)}_{\left( \frac{1}{4},\frac{1}{4},\frac{1}{2} \right)}\ 
{\cal L}\left[ 1,1 \right] ^2 {\cal L}\left[ 2,2 \right] 
\right. }  + \non 
& & \left. 16 
{\cal K}^{(1)}_{\left( \frac{1}{4},\frac{1}{4},-\frac{1}{2} \right)} 
+ 8 
{\cal K}^{(2)}_{\left( \frac{1}{4},-\frac{1}{4},0 \right)}\
{\cal L}\left[ 1,1 \right] 
+ 16 
{\cal K}^{(3)}_{\left( \frac{1}{4},\frac{1}{4},-\frac{1}{2} \right)} 
\right) ,
\eeqn
using the notations defined in the appendix. 
Numerically, all the twisted sector contributions vanish, 
but we have spelled them out to demonstrate the formal appearance of 
the projector in the tree channel. The modular $S$ transformation leads to 
\beqn
\tilde{\cal K} &=& 32c\ \int_0^\infty{dl\ 
\left( 
\tilde{\cal K}^{(0)}_{\left( \frac{1}{4},\frac{1}{4},-\frac{1}{2} \right)}\ 
\tilde{\cal L}\left[ 4,4 \right] ^2 \tilde{\cal L}\left[ 2,2 \right] 
\right. }  + \non 
& & \left. 4 
\tilde{\cal K}^{(1)}_{\left( \frac{1}{4},\frac{1}{4},-\frac{1}{2} \right)} 
- 4
\tilde{\cal K}^{(2)}_{\left( \frac{1}{4},-\frac{1}{4},0 \right)}\
\tilde{\cal L}\left[ 4,4 \right] 
- 4 
\tilde{\cal K}^{(3)}_{\left( \frac{1}{4},\frac{1}{4},-\frac{1}{2} \right)} 
\right) .
\eeqn
The relative factors are in perfect match with the expected ones
in (\ref{pref}).  This is also true
for the {\bf ABB} lattice, as the only difference is an overall factor
of $1/2$, originating both from the lattice 
partition functions and from the reduction of the number of invariant 
fixed points. This is quite a remarkable coincidence, indeed. \\

\begin{figure}[h]
\begin{center}
\makebox[10cm]{
 \epsfxsize=12cm
 \epsfysize=6cm
 \epsfbox{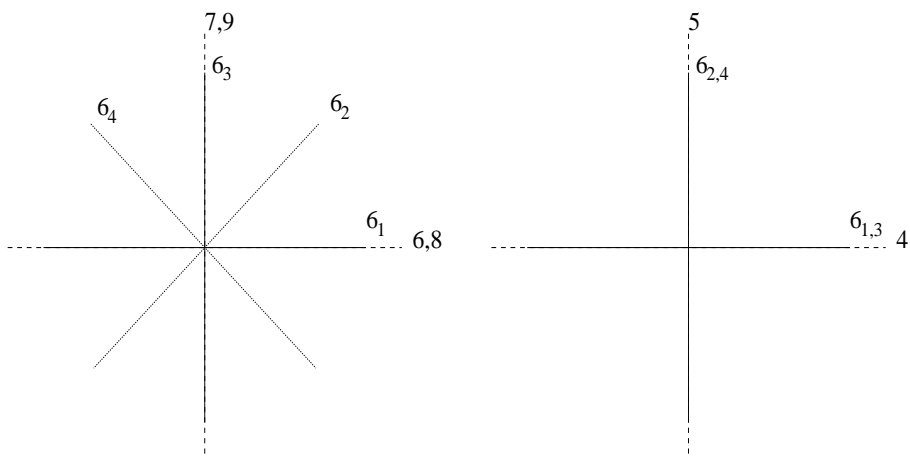}
}
\end{center}
\caption{Branes on the $\mathbb{Z}_4$ torus}
\label{figz4branes}
\end{figure}

To cancel the tadpole of the Klein bottle, we introduce D6$_i$-branes into the 
background as shown in Figure \ref{figz4branes}, where $i=1,\ldots,4$ mod $4$.
The relative angles are $\pi v_i$ and the $6_1$-branes are lying 
entirely inside the fixed plane of ${\cal R}$. 
%%%%%%%%%%%%%%%%%%%%%%%%%%%%%%%%%%%%%%%%%%%%%%%%%%%%%
%   BILD: 
%%%%%%%%%%%%%%%%%%%%%%%%%%%%%%%%%%%%%%%%%%%%%%%%%%%%%
%%%%%%%%%%%%%%%%%%%%%%%%%%%%%%%%%%%%%%%%%%%%%%%%%%%%%
%
%%%%%%%%%%%%%%%%%%%%%%%%%%%%%%%%%%%%%%%%%%%%%%%%%%%%%

As the orientifold group does not mix even and odd numbered branes, they will 
carry distinct factors of the gauge group and the only fields that 
transform non-trivially under both factors arise in the 
$\left( 6_i,6_{i+1} \right)$ and $\left( 6_i,6_{i+3} \right)$ open string 
sectors. 

As we have already anticipated in section 2, putting for instance 
the D6-branes in the third $T^2$  at an angle $\phi=\pi /4$ relative to 
the $x_4$ axes we would get  both different KK and W contributions 
and different intersection numbers. The transformation to tree channel
shows that for this choice of the D6-branes there is no chance
to cancel tadpoles. Thus, tadpole cancellation uniquely fixes the 
location of the D6-branes. 

For the correct location of the D6-branes shown in Figure \ref{figz4branes}
the complete annulus amplitude reads
\beqn
\label{annuloop}
{\cal A} &=& \frac{c}{4}\ \int_0^\infty{\frac{dt}{t^3}\ 
\left( 
{\rm M}^2 
{\cal A}^{(0,0)}_{\left( \frac{1}{4},\frac{1}{4},-\frac{1}{2} \right)}\ 
{\cal L}\left[ 2,2 \right] ^2 {\cal L}\left[ 1,1 \right] 
+ {1\over 4}\left( \sum_{i=1}^4 \gamma_i^2 \right) 
{\cal A}^{(0,2)}_{\left( \frac{1}{4},-\frac{1}{4},0 \right)}\ 
{\cal L}\left[ 2,2 \right] 
+ \right. } \non 
& & \left. 
{\rm M}^2 
{\cal A}^{(1,0)}_{\left( \frac{1}{4},\frac{1}{4},-\frac{1}{2} \right)}
+ {1\over 4}\left( \sum_{i=1}^4 \gamma_i\gamma_{i+1} \right) 
{\cal A}^{(1,2)}_{\left( \frac{1}{4},\frac{1}{4},-\frac{1}{2} \right)}
+ 2{\rm M}^2 
{\cal A}^{(2,0)}_{\left( \frac{1}{4},-\frac{1}{4},0 \right)}\ 
{\cal L}\left[ 2,2 \right] 
+ \right. \\ 
& & \left.
\left( \gamma_{1}\gamma_3+\gamma_2\gamma_4 \right) 
{\cal A}^{(2,2)}_{\left( \frac{1}{4},-\frac{1}{4},0 \right)}\ 
{\cal L}\left[ 2,2 \right] 
+ {\rm M}^2 
{\cal A}^{(3,0)}_{\left( \frac{1}{4},\frac{1}{4},-\frac{1}{2} \right)}
+  {1\over 4}\left( \sum_{i=1}^4 \gamma_i\gamma_{i+1} \right)
{\cal A}^{(3,2)}_{\left( \frac{1}{4},\frac{1}{4},-\frac{1}{2} \right)}
\right) \nonumber ,
\eeqn
where we have introduced the notation 
$\gamma_i={\rm tr}\left(\gamma_2^{(i)}\right)$  
for the action of the $\mathbb{Z}_2$ element on the Chan-Paton 
factors for the four different kinds of D6$_i$-branes. 
The amplitude (\ref{annuloop}) transforms to the tree channel expression:
\beqn
\tilde{\cal A} &=& \frac{c}{8} \int_0^\infty{dl\ 
\left( 
{\rm M}^2 
\tilde{\cal A}^{(0,0)}_{\left( \frac{1}{4},\frac{1}{4},-\frac{1}{2} \right)}\ 
\tilde{\cal L}\left[ 1,1 \right] ^2 \tilde{\cal L}\left[ 2,2 \right] 
+ 2 \left( \sum_{i=1}^4 \gamma_i^2 \right)
\tilde{\cal A}^{(0,2)}_{\left( \frac{1}{4},-\frac{1}{4},0 \right)}\ 
\tilde{\cal L}\left[ 1,1 \right] 
+ \right. } \non 
& & \left. 
4 {\rm M}^2 
\tilde{\cal A}^{(1,0)}_{\left( \frac{1}{4},\frac{1}{4},-\frac{1}{2} \right)}
+ \left( \sum_{i=1}^4 \gamma_i\gamma_{i+1} \right) 
\tilde{\cal A}^{(1,2)}_{\left( \frac{1}{4},\frac{1}{4},-\frac{1}{2} \right)}
- 4 {\rm M}^2 
\tilde{\cal A}^{(2,0)}_{\left( \frac{1}{4},-\frac{1}{4},0 \right)}\ 
\tilde{\cal L}\left[ 1,1 \right] 
+ \right. \\
& & \left.
2 \left( \gamma_{1}\gamma_3+\gamma_2\gamma_4 \right)
\tilde{\cal A}^{(2,2)}_{\left( \frac{1}{4},-\frac{1}{4},0 \right)}\ 
\tilde{\cal L}\left[ 1,1 \right] 
- 4 {\rm M}^2 
\tilde{\cal A}^{(3,0)}_{\left( \frac{1}{4},\frac{1}{4},-\frac{1}{2} \right)}
- \left( \sum_{i=1}^4 \gamma_i\gamma_{i+1} \right) 
\tilde{\cal A}^{(3,2)}_{\left( \frac{1}{4},\frac{1}{4},-\frac{1}{2} \right)}
\right) \nonumber .
\eeqn
If we change the lattice to {\bf ABB}, while keeping the branes fixed, 
we get an extra overall factor of 2 due to
different KK and W sums as well as a different intersection number for
$(6_i,6_{i+1})$ strings. In both cases the projector 
in the tree channel is complete. \\

The final contribution comes from the M\"obius strip amplitude, where 
the invariant windings and momenta are independent of the 
particular choice of the lattice: 
\beqn
\label{moebiloop}
{\cal M} &=& -\frac{c\, {\rm M}}{4} \int_0^\infty{\frac{dt}{t^3}\ 
\left( 
{\cal M}^{(0,0)}_{\left( \frac{1}{4},\frac{1}{4},-\frac{1}{2} \right)}\
{\cal L} \left[ 2,2 \right]^2 {\cal L}\left[ 1,4 \right]  \right. } \non
& & \left. 
-4 
{\cal M}^{(2,1)}_{\left( \frac{1}{4},\frac{1}{4},-\frac{1}{2} \right)}+
{\cal M}^{(0,2)}_{\left( \frac{1}{4},-\frac{1}{4},0 \right)}\
{\cal L} \left[ 2,2 \right]  
+4 {\cal M}^{(2,3)}_{\left( \frac{1}{4},\frac{1}{4},-\frac{1}{2} \right)}
\right) .
\eeqn
This leads to the tree channel expression
\beqn
\tilde{\cal M} &=& -4c\, {\rm M} \int_0^\infty{dl\ 
\left( 
\tilde{\cal M}^{(0)}_{\left( \frac{1}{4},\frac{1}{4},-\frac{1}{2} \right)}\
\tilde{\cal L} \left[ 4,4 \right]^2 \tilde{\cal L}\left[ 8,2 \right]  
\right. } \non
& & \left. 
-4 \tilde{\cal M}^{(3)}_{\left( \frac{1}{4},\frac{1}{4},-\frac{1}{2} \right)}-
4 \tilde{\cal M}^{(2)}_{\left( \frac{1}{4},-\frac{1}{4},0 \right)}\
\tilde{\cal L} \left[ 4,4 \right]
+4
\tilde{\cal M}^{(1)}_{\left( \frac{1}{4},\frac{1}{4},-\frac{1}{2} \right)}
\right) .
\eeqn
There are no extra factors whatsoever in the M\"obius strip 
amplitude when switching to the {\bf ABB} lattice. 
Summarizing, we get the following untwisted and $\mathbb{Z}_2$ twisted
tadpole cancellation conditions for the {\bf ABA} model
\beqn
\frac{1}{8} \left( {\rm M}^2 -32 {\rm M} + 256 \right) &=& 
\frac{1}{8} \left( {\rm M}-16 \right)^2 = 0,\non
\gamma_1^2+\gamma_3^2+\gamma_1\gamma_3 &=& 0, \non
\gamma_2^2+\gamma_4^2+\gamma_2\gamma_4 &=& 0 .
\eeqn
Deriving the $\mathbb{Z}_2$ twisted tadpole cancellation conditions one 
realizes that the $6_{1,3}$ and $6_{2,4}$ branes
are charged under different twisted RR 5-forms. This can be seen by
taking all the volume factors appearing in the amplitude
into account individually. 
More precisely, the twisted sector tadpole condition 
receives contributions from the
sixteen individual $\mathbb{Z}_2$ fixed points. Analyzing the intersection
of the D6-branes
with the 16 fixed points allows us to write each twisted sector tadpole 
condition as a sum of 6 perfect squares
\beqn 
\label{tadtw}
2\times\left[\left(\gamma_1+\gamma_3\right)^2+
      \gamma_1^2+\gamma_3^2\right] &=& 0, \non
2\times\left[\left(\gamma_2+\gamma_4\right)^2+
      \gamma_2^2+\gamma_4^2\right] &=& 0.
\eeqn
Apparently, in order to satisfy these 12 conditions we have to choose
$\gamma_i=0$ for all $i\in\{1,2,3,4\}$. 

For the {\bf ABB} configuration the untwisted tadpole condition becomes
\beqn
\frac{1}{4} \left( {\rm M}^2 -16 {\rm M} + 64 \right) = 
\frac{1}{4} \left( {\rm M}-8 \right)^2 =0 
\eeqn
with unchanged conditions for the twisted sector tadpoles.
Thus, the  $\mathbb{Z}_4$ orientifold gives rise to two different 
models, one with gauge group of rank 16 and the other
one with gauge group of rank 8.  \\

The massless closed string spectrum of the {\bf ABA} model 
receives six chiral multiplets from the untwisted sector besides 
the graviton and dilaton multiplets. 
Moreover, in the $\Theta^{1,3}$ twisted sectors the 16 fixed points
are all invariant under ${\cal R}$ and give rise to  16 chiral multiplets.
Finally, the 16 fixed points of $\Theta^2$
need to be distinguished with respect to their mapping under $\Theta, {\cal R}$ and
even $\Theta {\cal R}$ providing  another 15 chiral and 1 vector multiplets. 

Open strings in the $(6_{i},6_{i})$ sector
carry eight massless states, 
which provide a vector multiplet in the gauge group $U(8)\times U(8)$,
a chiral multiplet in the adjoint
representation $\left( ({\bf 64,1}) 
\oplus ({\bf 1, 64}) \right)$ and
two chiral multiplets in the
$ \left( ({\bf 28,1}) \oplus ({\bf \overline{28}, 1 }) \oplus
({\bf 1, 28}) \oplus ({\bf 1,\overline{28} })  \right)$ representation.  
In the $(6_{i},6_{i+2})$ sector one has to take into account the extra
minus signs in the loop channel M\"obius strip amplitude (\ref{moebiloop}),
implying that compared to the  $(6_{i},6_{i})$ sector both the
$\Omega{\cal R}$ and the $\Theta^2$ projection change sign.
Considering also the multiplicity due to the twofold degeneracy of the
ground state in this sector and an extra factor of two from 
the intersections between $6_{i}$ and $6_{i+2}$ branes
one gets two chiral multiplets in the $\left( ({\bf 64,1}) 
\oplus ({\bf 1, 64}) \right)$ representation. 
Finally the $(6_{i},6_{i+1})$ strings only carry a single massless state 
to be counted with two orientations
giving rise to one chiral multiplet in the 
$\left( {\bf (8,\overline{8})} \oplus {\bf (\overline{8},{8})} \right)$
representation. 
Inspecting the massless open string spectrum reveals that it is not only
non-chiral but surprisingly fits into ${\cal N}=2$ multiplets.
However, the D-branes in the $(6_{i},6_{i+1})$ sector only preserve
${\cal N}=1$ supersymmetry, so that the appearance of ${\cal N}=2$ multiplets
is purely accidental. 
The complete ${\cal N}=1$ supersymmetric massless 
spectrum is shown in the Tables \ref{tableclosed} and \ref{tableopenz4}, 
where we have also summarized the results for the ${\bf ABB}$ model. \\

\begin{table}[h]
\caption{${\cal N}=1$ open spectra for the $\mathbb{Z}_4$}
\label{tableopenz4}
\begin{center} 
\begin{tabular}{|l|l|c|c|}
\hline 
Model & $\left( 6_{i},6_{i} \right)$ & 
        $\left( 6_{i},6_{i+1} \right)$  & 
        $\left( 6_{i},6_{i+2} \right)$  \\ 
\hline 
${\bf ABA}$ & $U(8) \times U(8)$ & 
        ${\bf (8,\overline{8})} \oplus {\bf (\overline{8},8)}$ & 
        $2{\bf (64,1)} \oplus 2{\bf (1,64)}$ \\
& ${\bf (64,1)} \oplus {\bf (1,64)}\oplus$ & & \\ 
& $2{\bf (28,1)} \oplus 2{\bf (\overline{28},1)}\oplus$ & & \\
& $2{\bf (1,28)} \oplus 2{\bf (1,\overline{28})}$ & & \\
\hline
${\bf ABB}$ & $U(4) \times U(4)$ & 
        $2{\bf (4,\overline{4})} \oplus 2{\bf (\overline{4},4)}$ & 
        $2{\bf (16,1)} \oplus 2{\bf (1,16)}$ \\
& ${\bf (16,1)} \oplus {\bf (1,16)}\oplus$ & & \\ 
& $2{\bf (6,1)} \oplus 2{\bf (\overline{6},1)}\oplus$ & & \\
& $2{\bf (1,6)} \oplus 2{\bf (1,\overline{6})}$ & & \\
\hline
\end{tabular}
\end{center}
\end{table}

\section{Results for the $\mathbb{Z}_3$ orientifold}

In this section we make some remarks on the specialities of the 
other models we have explicitly computed. 
We show how the lattices and brane configurations look 
like, give the tadpole cancellation conditions and their solutions 
in terms of the gauge groups. 
Finally we display the massless open and closed string 
spectra. The details of the computation are collected in the appendix. \\

%%%%%%%%%%%%%%%%%%%%%%%%%%%%%%%%%%%%%%%%%%%%%%%%%%%%%
%   BILD: 
%%%%%%%%%%%%%%%%%%%%%%%%%%%%%%%%%%%%%%%%%%%%%%%%%%%%%
\begin{figure}[h]
\begin{center}
\makebox[10cm]{
 \epsfxsize=6cm
 \epsfysize=6cm
 \epsfbox{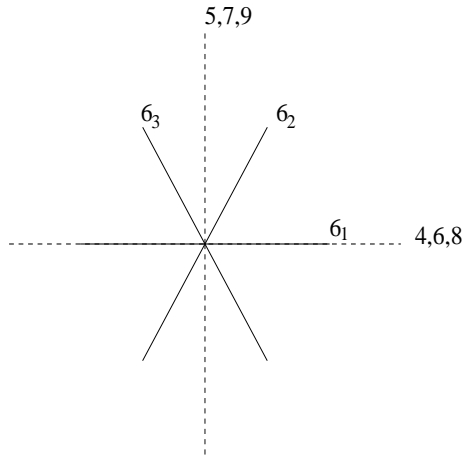}
}
\end{center}
\caption{Branes on the $\mathbb{Z}_3$ torus}
\label{figz3branes}
\end{figure}
%%%%%%%%%%%%%%%%%%%%%%%%%%%%%%%%%%%%%%%%%%%%%%%%%%%%%
%
%%%%%%%%%%%%%%%%%%%%%%%%%%%%%%%%%%%%%%%%%%%%%%%%%%%%%

The $\mathbb{Z}_3$ orientifold is very similar to the 
six-dimensional $\mathbb{Z}_3$ model that has been discussed in some detail in 
\cite{hepth9908130}. As was pointed out in \cite{AngBlu} 
one is free to choose any of the  four possible lattices
$\{{\bf AAA},{\bf AAB},{\bf ABB},{\bf BBB}\}$. 
In all four cases the D6-branes are located in the same way on all three 
2-tori as shown in Figure \ref{figz3branes}. 
The lattice ${\bf A}^{3-i}{\bf B}^i$ on the one hand leads
to an overall factor of $3^i$ for 
all the amplitudes and on the other hand causes
the number of intersections to be  $3^i$, too. 
These extra factors affect  the multiplicities for
the different massless open string modes. 
In all cases the tadpole cancellation condition reads
\beqn
\frac{1}{2} \left( {\rm M}^2 -8 {\rm M} + 16 \right) = 
\frac{1}{2} \left( {\rm M}-4 \right)^2 =0, 
\eeqn  
giving an $SO(4)$ gauge symmetry. 
The ${\cal N}=1$ supersymmetric spectra of all the 
models are collected in the Tables \ref{tableclosed} and \ref{tableopenz3}. \\

\begin{table}[h]
\caption{${\cal N}=1$ open spectra for the $\mathbb{Z}_3$}
\label{tableopenz3}
\begin{center} 
\begin{tabular}{|l|l|c|}
\hline 
Model & $\left( 6_{i},6_{i} \right)$ & 
        $\left( 6_{i},6_{i+1} \right)$  \\
\hline 
${\bf AAA}$ & $SO(4)+3({\bf 6})$ & $({\bf 10})$ \\
${\bf AAB}$ & $SO(4)+3({\bf 6})$ & $3({\bf 10})$ \\
${\bf ABB}$ & $SO(4)+3({\bf 6})$ & $9({\bf 10})$ \\
${\bf BBB}$ & $SO(4)+3({\bf 6})$ & $27({\bf 10})$ \\
\hline 
\end{tabular}
\end{center}
\end{table}

\noindent Note, that
in the  $\left( 6_{i},6_{i+1} \right)$ sector, the phases in the 
definition and modular properties of the
$\vartheta$-functions  lead to the symmetric representation
of the gauge group.\\ 

\clearpage

\section{Results for the $\mathbb{Z}_6$ and $\mathbb{Z}_6^\prime$ 
orientifolds} 

\begin{figure}[h]
\begin{center}
\makebox[10cm]{
 \epsfxsize=12cm
 \epsfysize=7cm
 \epsfbox{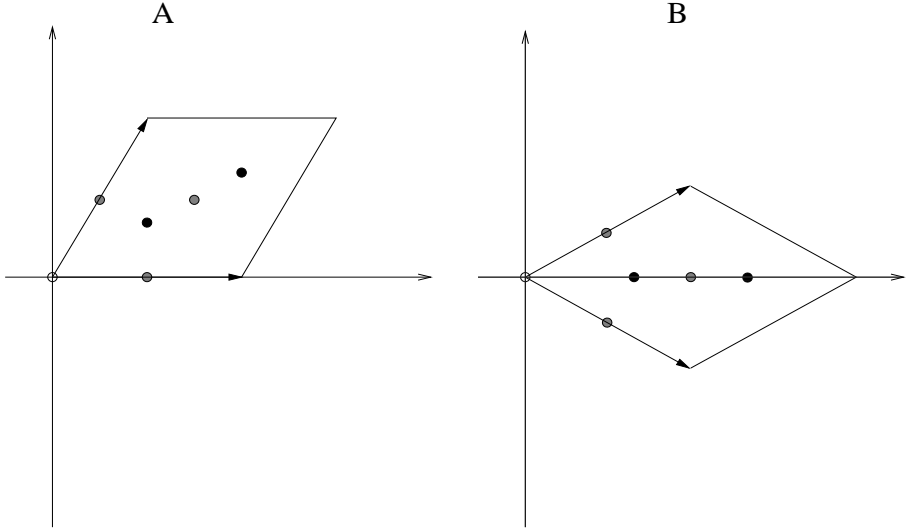}
}
\end{center}
\caption{The $\mathbb{Z}_6$ lattices}
\label{figz6zellen}
\end{figure}

For each of the two $\mathbb{Z}_6$ orbifold groups we obtain two inequivalent 
models on the massless level. 
As mentioned in section 2, we are forced to choose lattices ${\bf AB}$ or
${\bf BA}$ in the two directions with $Nv_i$ odd
and can arbitrarily choose the lattice for the third $T^2$  to be 
either ${\bf A}$ or ${\bf B}$.
The two lattices of ${\bf A}$ and  ${\bf B}$ type are shown 
in Figure \ref{figz6zellen}. \\
%%%%%%%%%%%%%%%%%%%%%%%%%%%%%%%%%%%%%%%%%%%%%%%%%%%%%
%   BILD: 
%%%%%%%%%%%%%%%%%%%%%%%%%%%%%%%%%%%%%%%%%%%%%%%%%%%%%
%%%%%%%%%%%%%%%%%%%%%%%%%%%%%%%%%%%%%%%%%%%%%%%%%%%%%
%
%%%%%%%%%%%%%%%%%%%%%%%%%%%%%%%%%%%%%%%%%%%%%%%%%%%%%

\begin{figure}[h]
\begin{center}
\makebox[10cm]{
 \epsfxsize=10cm
 \epsfysize=5cm
 \epsfbox{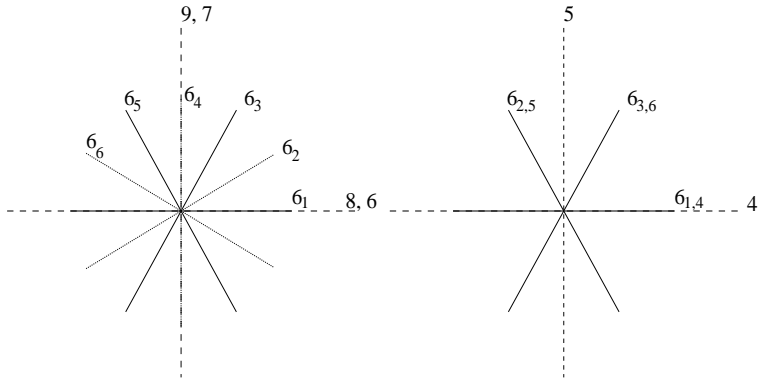}
}
\end{center}
\caption{Branes on the $\mathbb{Z}_6$ torus}
\label{figz6branes}
\end{figure}

\noindent
The location  of the D6-branes is chosen according to the general rules 
and we depict it for the $\mathbb{Z}_6$ orientifold in 
Figure \ref{figz6branes}. \\ 
%%%%%%%%%%%%%%%%%%%%%%%%%%%%%%%%%%%%%%%%%%%%%%%%%%%%%
%   BILD: 
%%%%%%%%%%%%%%%%%%%%%%%%%%%%%%%%%%%%%%%%%%%%%%%%%%%%%
%%%%%%%%%%%%%%%%%%%%%%%%%%%%%%%%%%%%%%%%%%%%%%%%%%%%%
%
%%%%%%%%%%%%%%%%%%%%%%%%%%%%%%%%%%%%%%%%%%%%%%%%%%%%%

\noindent The untwisted tadpole
cancellation condition in this case is
\beqn
\frac{1}{2} \left( {\rm M}^2 -8 {\rm M} + 16 \right) = 
\frac{1}{2} \left( {\rm M}-4 \right)^2 =0  .
\eeqn  
The $\mathbb{Z}_2$ twisted tadpole condition is similar to the one
in (\ref{tadtw}) and can be satisfied by choosing traceless 
$\gamma_3^{(i)}$ matrices, leading to a $U(2)\times U(2)$ gauge group. 
Taking all twisted sectors
and intersection points into account we derive the ${\cal N}=1$ 
supersymmetric open string massless
spectrum shown in Table \ref{tableopenz6}. To distinguish $SU(2)$ singlets 
which are 
charged or neutral under the $U(1)$ we use the notation ${\bf 1}$ and 
${\bf 1_0}$.
The ${\bf 2}$ and ${\bf \overline{2} }$ are similarly distinguished by
their abelian $U(1)$ charges. \\
\begin{center}
\begin{table}[h]
\caption{${\cal N}=1$ open spectra for the $\mathbb{Z}_6$}
\label{tableopenz6}
\begin{center} 
\begin{tabular}{|l|l|c|c|c|c|}
\hline 
Model & $\left( 6_{i},6_{i} \right)$ & 
        $\left( 6_{i},6_{i+1} \right)$  & 
        $\left( 6_{i},6_{i+2} \right)$  & 
        $\left( 6_{i},6_{i+3} \right)$  \\  
\hline 
${\bf ABA}$ & $U(2) \times U(2)$ & 
        ${\bf (2,\overline{2})} \oplus {\bf (\overline{2},2)}$ & 
        ${\bf (3,1_0)} \oplus {\bf (\overline{3},1_0)}\oplus$ & 
        $4{\bf (2,\overline{2})} \oplus 4{\bf (\overline{2},2)}$ \\
& ${\bf (4,1_0)} \oplus {\bf (1_0,4)}\oplus$ & 
& ${\bf (1_0,3)} \oplus {\bf (1_0,\overline{3})} \oplus$ & \\
& $2{\bf (1,1_0)} \oplus 2{\bf (\overline{1},1_0)}\oplus$
& & $2{\bf (4,1_0)} \oplus 
        2{\bf (1_0,4)}$ & \\
& $2{\bf (1_0,1)} \oplus 2{\bf (1_0,\overline{1})}$ & &  &\\
\hline 
${\bf ABB}$ & $U(2) \times U(2)$ & 
        $3{\bf (2,\overline{2})} \oplus 3{\bf (\overline{2},2)}$ & 
        $3{\bf (3,1_0)} \oplus 3{\bf (\overline{3},1_0)}\oplus$ & 
        $4{\bf (2,\overline{2})} \oplus 4{\bf (\overline{2},2)}$ \\
& ${\bf (4,1_0)} \oplus {\bf (1_0,4)}\oplus$ & 
& $3{\bf (1_0,3)} \oplus 3{\bf (1_0,\overline{3})} \oplus$ & \\
& $2{\bf (1,1_0)} \oplus 2{\bf (\overline{1},1_0)}\oplus$
& & $6{\bf (4,1_0)} \oplus 
        6{\bf (1_0,4)}$ & \\
& $2{\bf (1_0,1)} \oplus 2{\bf (1_0,\overline{1})}$ & &  &\\
\hline
\end{tabular}
\end{center}
\end{table}
\end{center}

\begin{figure}[h]
\begin{center}
\makebox[10cm]{
 \epsfxsize=15cm
 \epsfysize=5cm
 \epsfbox{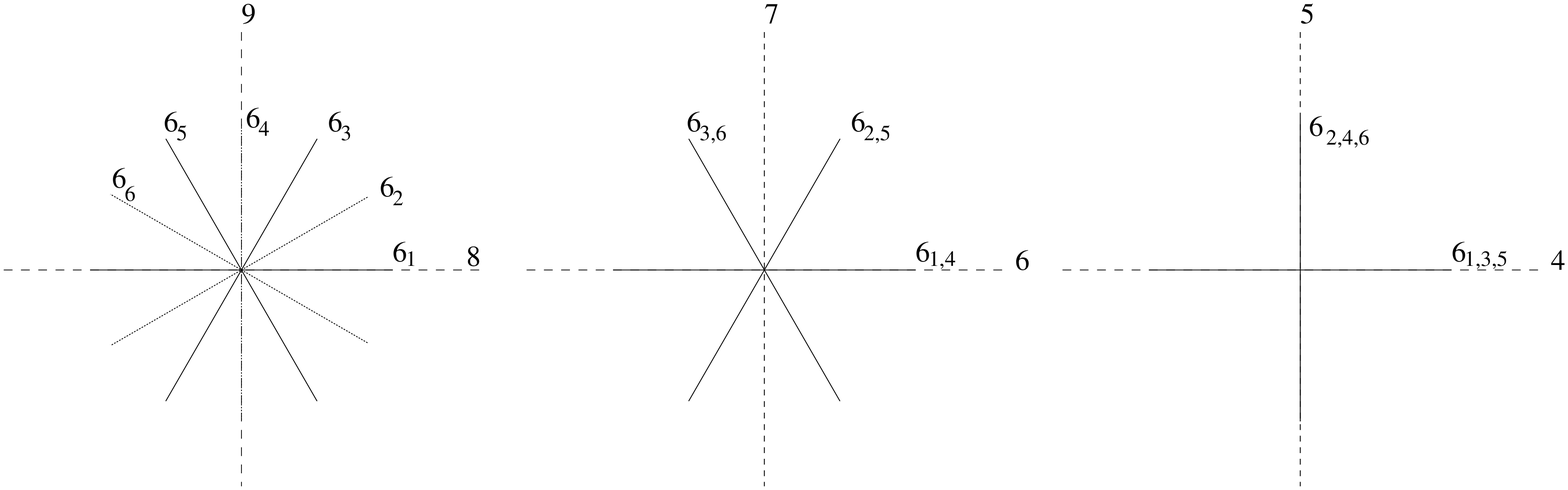}
}
\end{center}
\caption{Branes on the $\mathbb{Z}_6^\prime$ torus}
\label{figz6pbranes}
\end{figure}

\noindent
Finally we discuss the $\mathbb{Z}_6^\prime$ orientifold where 
one has to choose the D6-branes
as shown in Figure \ref{figz6pbranes}.  \\
%%%%%%%%%%%%%%%%%%%%%%%%%%%%%%%%%%%%%%%%%%%%%%%%%%%%%
%%%%%%%%%%%%%%%%%%%%%%%%%%%%%%%%%%%%%%%%%%%%%%%%%%%%%
%   BILD: 
%%%%%%%%%%%%%%%%%%%%%%%%%%%%%%%%%%%%%%%%%%%%%%%%%%%%%
%%%%%%%%%%%%%%%%%%%%%%%%%%%%%%%%%%%%%%%%%%%%%%%%%%%%%

\noindent
Special care needs to be taken in the $\Theta^2$-twisted sector of the 
$\mathbb{Z}_6^\prime$ model, as the partition function of the KK and W states 
as well as the number of fixed points and intersections does depend on 
the factor of the gauge group. Eventually, everything comes out just right
and leads to the complete projector in the tree channel. 
As a consequence the two choices ${\bf AB}$ and ${\bf BA}$ 
in the directions of odd $Nv_i$ lead to slightly different models. 
As can be seen by direct computation, on the massless level this 
difference is only an exchange of the two gauge
factors. 
 
Again the untwisted tadpole cancellation condition is
unaffected by the change and reads explicitly: 
\beqn
\frac{1}{2} \left( {\rm M}^2 -8 {\rm M} + 16 \right) = 
\frac{1}{2} \left( {\rm M}-4 \right)^2 =0.  
\eeqn  
The gauge group again is 
$U(2) \times U(2)$ and the ${\cal N}=1$ supersymmetric massless open string 
spectrum is shown in  Table \ref{tableopenz6p}. \\

\begin{center}
\begin{table}[h]
\caption{${\cal N}=1$ open spectra for the $\mathbb{Z}_6^\prime$}
\label{tableopenz6p}
\begin{center} 
\begin{tabular}{|l|l|c|c|c|c|}
\hline 
Model & $\left( 6_{i},6_{i} \right)$ & 
        $\left( 6_{i},6_{i+1} \right)$  & 
        $\left( 6_{i},6_{i+2} \right)$  & 
        $\left( 6_{i},6_{i+3} \right)$  \\  
\hline 
${\bf AAB}$ & $U(2) \times U(2)$ &
        $2{\bf (2,\overline{2})} \oplus 2{\bf (\overline{2},2)}$ & 
        ${\bf (1,1_0)} \oplus {\bf (\overline{1},1_0)}\oplus$ & 
        $4{\bf (2,\overline{2})} \oplus 4{\bf (\overline{2},2)}$ \\
& ${\bf (4,1_0)} \oplus {\bf (1_0,4)}\oplus$ & 
& $3{\bf (1_0,1)} \oplus 3{\bf (1_0,\overline{1})} \oplus$ & \\
& $2{\bf (1,1_0)} \oplus 2{\bf (\overline{1},1_0)}\oplus$
& & ${\bf (4,1_0)} \oplus 
        3{\bf (1_0,4)}$ & \\
& $2{\bf (1_0,1)} \oplus 2{\bf (1_0,\overline{1})}$ & &  &\\
\hline
${\bf ABB}$ & $U(2) \times U(2)$ & 
        $6{\bf (2,\overline{2})} \oplus 6{\bf (\overline{2},2)}$ & 
        $3{\bf (1,1_0)} \oplus 3{\bf (\overline{1},1_0)}\oplus$ & 
        $4{\bf (2,\overline{2})} \oplus 4{\bf (\overline{2},2)}$ \\
& ${\bf (4,1_0)} \oplus {\bf (1_0,4)}\oplus$ & 
& $9{\bf (1_0,1)} \oplus 9{\bf (1_0,\overline{1})} \oplus$ & \\
& $2{\bf (1,1_0)} \oplus 2{\bf (\overline{1},1_0)}\oplus$
& & $3{\bf (4,1_0)} \oplus 
        9{\bf (1_0,4)}$ & \\
& $2{\bf (1_0,1)} \oplus 2{\bf (1_0,\overline{1})}$ & &  &\\
\hline
\end{tabular}
\end{center}
\end{table}
\end{center}

\noindent
We used the first gauge factor for the theory on the odd $D6_i$-branes. 
In contrast to the other cases discussed so far,  the massless spectrum of the 
$\mathbb{Z}_6^\prime$ orientifold is 
not invariant under exchange of the two $U(2)$ gauge factors. 

\section{Conclusions}

In this paper we have studied $\Omega{\cal R}$ orientifolds of type IIA
in four dimensions. We found a couple of models for which the tadpole
cancellation conditions could be satisfied, leading to non-chiral
massless spectra on the D6-branes. 
It would be interesting to
study the other ${\cal N}=1$ supersymmetric orbifolds from Table 1, as
well. However, for these cases the contribution of the lattices is more
involved, as they cannot be factorized as $T^6=T^2\times T^2\times T^2$.
Even knowing the basis vectors of $T^6$ tori with $\mathbb{Z}_7$,
$\mathbb{Z}_8$ and $\mathbb{Z}_{12}$ symmetry is not enough, one also
has to orientate the lattice in such a way that also the reflection
${\cal R}$ acts crystallograhically, which may give rise to a variety of 
distinct models.

As we pointed out in \cite{hepth9908130} these $\Omega{\cal R}$ orientifolds 
might also be interesting for non-supersymmetric generalizations.
In particular, the $\Omega (-1)^{F_L}$ 
orientifolds of type 0B are known to be sometimes free of tachyons
leading to classically stable non-supersymmetric string vacua \cite{rnonsusy}.
However, for standard $\mathbb{Z}_N$ orientifolds with $N\ge 3$
some tachyons appearing in the twisted sectors
survive the $\Omega$ projection. The reason is, that $\Omega$
exchanges the $\Theta^k$ twisted sector with the $\Theta^{-k}$ twisted sector,
so that one linear combination of both tachyons survives. 
Since in the $\Omega{\cal R}$ case, the twisted sectors are left invariant
one may hope that at least the closed string sector of type 0A $\mathbb{Z}_N$
orientifolds is free of tachyons. Sometimes this is indeed the case.
However, it is easy to see that the tachyons now return in the 
open string sectors, namely in the twisted open string sectors.
Thus, in contrast to our earlier expectation type 0A 
$\Omega{\cal R}(-1)^{F_L}$ 
orientifolds also generically contain open string tachyons and do not
behave  better than standard type 0B $\Omega (-1)^{F_L}$ orientifolds. \\ 

\vspace{1cm}

\noindent {\bf Acknowledgements} \\ 

We would like to thank A. Kumar, who was involved in early stages of 
this work, as well as M. Gaberdiel, D. L\"ust, A. Miemiec and 
H. Skarke for encouraging discussions and helpful remarks.  

\clearpage
\begin{appendix}

\section{One-loop partition functions}

The general strategy to compute the several one loop contributions to the 
massless tadpoles is as usual: We compute the loop diagrams of the 
Klein bottle, the annulus and the M\"obius strip and convert the results 
via a modular $S$ transformation sending $t \mapsto 1/t$ into the 
tree channel, expand the integrand for large distances and extract the 
coefficients of the divergent terms. This gives us a condition for the number 
of branes and for the action of non-trivially represented Chan-Paton 
matrices, the only one being $\gamma_{N/2}^{(i)}$ for the $\mathbb{Z}_N$ 
orbifolds with even $N$. In general we denote the unitary action on 
Chan-Paton labels by $\gamma_k^{(i)}$ 
and $\gamma_{\Omega{\cal R}k}^{(i)}$ for $\Theta^k$ and 
$\Omega{\cal R}\Theta^k$ acting on the D6$_i$-branes respectively.

\subsection{Klein bottle}

The Klein bottle amplitude in the loop channel has the general form
\beqn \label{KBpafu}
{\cal K} = 4\frac{V_4}{(8 \pi^2 \al )^2} \int_0^\infty{\frac{dt}{t^3} 
\left(\frac{1}{4N}  \sum_{n,k=0}^N{{\cal K}^{(n,k)}\ 
{\cal L}_{\cal K}^{(n,k)}  
} \right) } 
\eeqn 
with ${\cal K}^{(n,k)}$ denoting the trace over oscillators (osc)
in the sector 
twisted by $\Theta^n$ with the insertion of $\Theta^k$ inside
\beqn
{\cal K}^{(n,k)} \equiv \Tr_{\mbox{ \sz osc}}^{(n)} 
\left( \Omega{\cal R}\Theta^k\  
        \left( 1+(-1)^F \right) 
        e^{-2\pi t \left( L_0 + \bar{L}_0 \right)} 
\right) 
\eeqn
and ${\cal L}_{\cal K}^{(n,k)}$ 
standing for the trace over bosonic zero modes, 
quantized momenta and windings (KK+W), i.e. 
\beqn
{\cal L}_{\cal K}^{(n,k)} \equiv \chi_{{\cal K}}^{(n,k)}
\Tr_{\mbox{ \sz KK+W}}^{(n)} 
\left( \Omega{\cal R}\Theta^k\  
        e^{-2\pi t \left( L_0 + \bar{L}_0 \right)}  
\right) .
\eeqn
Here $\chi_{{\cal K}}^{(n,k)}$ is the number of 
fixed points of $\Theta^n$, which 
are invariant under the operator ${\cal R} \Theta^k$ in the trace. 
Except for the $\mathbb{Z}_6^\prime$ orbifold these numbers, as well 
as the entire traces are equal 
for any insertion of $\Theta^k$ and even in that case only the lattice 
contributions differ, so we can factorize the partition function as 
in (\ref{KBpafu}). In all other cases we omit the superscript $k$, 
writing ${\cal L}_{\cal K}^{(n)}$, ${\cal K}^{(n)}$ and 
$\chi_{{\cal K}}^{(n)}$. 
As a shorthand for the contributions of the bosonic zero modes we use
\beqn \label{lattpafu}
{\cal L} \left[ \alpha, \beta \right] \equiv  
\left( \sum_{m\in \mathbb{Z}}{e^{-\alpha \pi t m^2/R^2} }\right)
\left( \sum_{n\in \mathbb{Z}}{e^{-\beta \pi t n^2 R^2} }\right) .
\eeqn
For all amplitude we used the convention
\beqn
\sum_{(nv_i,kv_i) \notin \mathbb{Z}^2} nv_i =0.
\eeqn
The oscillator sums lead to the generic expression
\beqn
{\cal K}^{(n)}_{\left( v_1,v_2,v_3 \right)} \equiv (1-1)\ 
\frac{\vartheta \left[ 0 \atop 1/2 \right]}{\eta^3} 
\prod_{nv_i \notin \mathbb{Z}}{ \left( 
        \frac{\vartheta \left[ nv_i \atop 1/2 \right]}
        {\vartheta \left[ 1/2+ nv_i \atop 1/2 \right]}
        e^{\pi i \langle nv_i \rangle}
        \right)
}
\prod_{nv_i \in \mathbb{Z}}{ \left(  
        \frac{\vartheta \left[ 0 \atop 1/2 \right]}{\eta^3} 
        \right)
} 
\eeqn
with argument $q= \exp \left( -4\pi t\right)$ and
\beqn
\langle nv_i \rangle\equiv nv_i-[nv_i]-{1\over 2}.
\eeqn
${\cal L}_{\cal K}^{(n,k)}$ has contributions whenever $nv_i \in \mathbb{Z}$. 
In contrast to the oscillator 
part we cannot give any generic formula for the lattice sums, since one has 
to take into account which momentum and winding states are invariant under 
the respective operator in the trace. This does not only depend on the 
orbifold group action, as given by the $v_i$, but also on the orientation 
of the lattice with respect to the reflection ${\cal R}$. 
For the tree channel amplitudes we also define appropriate abbreviations:
\beqn
\tilde{\cal K}^{(n)}_{\left( v_1,v_2,v_3 \right)} & \equiv & (1-1)\ 
\frac{\vartheta \left[ 1/2 \atop 0 \right]}{\eta^3} 
\prod_{nv_i \notin \mathbb{Z}}{ \left( 
        \frac{\vartheta \left[ 1/2 \atop nv_i \right]}
        {\vartheta \left[ 1/2 \atop 1/2+ nv_i \right]}
        \right)
}
\prod_{nv_i \in \mathbb{Z}}{ \left(  
        \frac{\vartheta \left[ 1/2 \atop 0 \right]}{\eta^3} 
        \right)
} , \non 
\tilde{\cal L} \left[ \alpha, \beta \right] & \equiv &  
\left( \sum_{m\in \mathbb{Z}}{e^{-\alpha \pi l m^2 R^2} }\right)
\left( \sum_{n\in \mathbb{Z}}{e^{-\beta \pi l n^2/R^2} }\right) 
\eeqn
the argument of $\tilde{\cal K}^{(n)}_{\left( v_1,v_2,v_3 \right)}$ being 
$\tilde{q} = \exp \left( -4\pi l\right)$. The expected prefactor to
yield the complete projector is therefore
\beqn
\prod_{nv_i \notin \mathbb{Z}} (-2 \sin (\pi nv_i)).
\eeqn
The volume factors from the modular 
transformation formally cancel out, the divergence being related to a 
tadpole of a 7-form field.

\subsection{Annulus}

The open string diagrams 
can also be factorized into oscillator and momentum and winding parts. 
For the annulus the loop channel reads  
\beqn \label{Apafu}
{\cal A} = \frac{V_4}{(8 \pi^2 \al )^2} \int_0^\infty \frac{dt}{t^3}  
    \left( \frac{1}{4N}
        \sum_{n,k,i=0}^N \tr \left( \gamma_k^{(i)} \right)
         \tr \left(  \left( \gamma_k^{(i+n)} \right)^{-1} \right)\ 
        {\cal A}^{(n,k)}\ {\cal L}_{\cal A}^{(n,k,i)} \right) 
\eeqn
and analogously to the above we define
\beqn
{\cal A}^{(n,k)} \equiv \Tr_{\mbox{ \sz osc}}^{(i,i+n)} 
\left( \Theta^k  
        \left( 1+(-1)^F \right)  
        e^{-2\pi t L_0} 
\right) ,
\eeqn
where the trace is to be performed over the oscillator excitations 
of the open strings stretching between the $i$ and $i+n$ branes, with 
only $k\in \{ 0,N/2\}$ contributions non-vanishing. Concerning the 
Chan-Paton matrices, we choose 
$\tr \left( \gamma_0^{(i)}\right)^2={\rm M}^2$, leaving only 
$\tr \left( \gamma_{N/2}^{(i)} \right)$ and the number of branes, 
${\rm M}$, to be determined by the tadpole 
cancellation conditions. The bosonic zero modes contribute  
\beqn 
\label{trkk}
{\cal L}_{\cal A}^{(n,k,\pm)} \equiv \chi_{\cal A}^{(n,k,\pm)}\    
\Tr_{\mbox{ \sz KK+W}}^{(i,i+n)} 
\left( \Theta^k\ 
        e^{-2\pi t L_0} 
\right) .
\eeqn
$\chi_{\cal A}^{(n,k,\pm)}$ is the number of intersections 
of the two types of branes on the torus, which are invariant under 
$\Theta^k$. The trace (\ref{trkk}) is only different from 1 if the $i$ 
and $i+n$ branes coincide in some (real) 
direction on the torus $T^6$ and $\Theta^k$ acts trivially 
in the complex plane of this direction. In the $\mathbb{Z}_6^\prime$ 
orbifold there is another distinction between odd and even $i$, 
which we have reserved the extra superscript $\pm$ for.   
For the oscillator part we get the generic formula
\beqn
{\cal A}^{(n,k)}_{\left( v_1,v_2,v_3\right)} \equiv (1-1)\ 
\frac{\vartheta \left[ 0 \atop 1/2 \right]}{\eta^3} 
\prod_{(nv_i,kv_i) \notin \mathbb{Z}^2}{ \left( 
        \frac{2^\delta \vartheta \left[ nv_i \atop 1/2+kv_i \right]}
        {\vartheta \left[ 1/2 + nv_i \atop 1/2 + kv_i \right]}
         e^{\pi i \langle nv_i\rangle}
        \right)
}
\prod_{(nv_i,kv_i) \in \mathbb{Z}^2}{ \left(  
        \frac{\vartheta \left[ 0 \atop 1/2 \right]}{\eta^3} 
        \right)
} 
\eeqn
with argument $q=\exp \left( -2\pi t\right)$. 
The second product is empty, except if both $nv_i$ and $kv_i$ are 
integers, which is also the only case, in which there are contributions 
to ${\cal L}^{(n,k,\pm)}_{\cal A}$. 
We have introduced factors of $2^\delta$ for cancelling 
inappropriate factors occurring in the theta functions by defining 
$\delta=1$ if $nv_i \in \mathbb{Z}$ and $kv_i \in \mathbb{Z}+1/2$ and 
$\delta=0$ otherwise. One could have omitted these extra factors by using 
the notations of \cite{riban}, for instance, but this would 
then imply a definition of the case $\delta=1$ via some limit of the 
otherwise undefined formula, which we preferred to avoid.  
Again there is no generic expression for the lattice sums 
and one needs to consider not only the type of brane as given by $i,n$ but 
also their orientation on the torus, in order to determine the normalization 
of momenta and winding states. ${\cal L}^{(n,k,\pm)}_{\cal A}$ can then be 
written in the form of (\ref{lattpafu}), of course.  
For the tree channel oscillators we use 
\beqn
\tilde{\cal A}^{(n,k)}_{\left( v_1,v_2,v_3\right)} \equiv (1-1)\ 
\frac{\vartheta \left[ 1/2 \atop 0 \right]}{\eta^3} 
\prod_{(nv_i,kv_i) \notin \mathbb{Z}^2}{ \left( 
        \frac{\vartheta \left[ 1/2+kv_i \atop nv_i \right]}
        {\vartheta \left[ 1/2 + kv_i \atop 1/2 + nv_i \right]}
        \right)
}
\prod_{(nv_i,kv_i) \in \mathbb{Z}^2}{ \left(  
        \frac{\vartheta \left[ 1/2 \atop 0 \right]}{\eta^3} 
        \right)
} 
\eeqn
with argument $\tilde{q}=\exp \left( -4\pi l\right)$. In the tree channel the 
annulus only contributes to the untwisted and $\Theta^{N/2}$ twisted sector, 
corresponding to the 7-form and a twisted 5-form tadpole.

\subsection{M\"obius strip}

Finally we need to go through the M\"obius strip amplitude. 
The loop channel expression is 
\beqn \label{Mpafu}
{\cal M} = - \frac{V_4}{(8 \pi^2 \al )^2} \int_0^\infty{\frac{dt}{t^3}  
\left(
        \frac{1}{4N}
        \sum_{n,k,i=0}^N{\tr \left( \left( \gamma_{\Omega{\cal R}k}^{(i)}\right)^{-1} 
        \left( \gamma_{\Omega{\cal R}k}^{(i)}\right)^{\mbox{\sz T}} \right)\ 
        {\cal M}^{(n,k)}\ 
        {\cal L}_{\cal M}^{(n,k,i)}}  
\right) } ,
\eeqn
with the oscillator 
\beqn
{\cal M}^{(n,k)} \equiv \Tr_{\mbox{ \sz osc}}^{(1,1+n)} 
\left( \Omega{\cal R}\Theta^k\  
        \left( 1+(-1)^F \right)  
        e^{-2\pi t L_0} 
\right) 
\eeqn
and the zero mode trace 
\beqn
{\cal L}_{\cal M}^{(n,k,\pm)} \equiv \chi_{\cal M}^{(n,k,\pm)}\    
\Tr_{\mbox{ \sz KK+W}}^{(i,i+n)} 
\left( \Omega{\cal R}\Theta^k\   
        e^{-2\pi t L_0} 
\right) .
\eeqn
Now $\chi_{\cal M}^{(n,k,\pm)}$ denotes the number of intersection points 
of the $i$ and $i+n$ branes, invariant under ${\cal R} \Theta^k$. By looking at the 
following chain of mappings of open strings  
\beqn
& & \left( i,i+n \right) \stackrel{\Theta^k}{\longrightarrow}
\left( i+2k,i+n+2k \right) \non 
& \stackrel{{\cal R}}{\longrightarrow} &
\left( 2-i-2k,2-i-n-2k \right) \stackrel{\Omega}{\longrightarrow}
\left( 2-i-n-2k,2-i-2k \right)  
\eeqn 
one realizes that only strings that satisfy $2(k+i-1)+n=0\ {\mbox{mod}}\ N$ 
can contribute in the M\"obius strip. If $N$ is even, the relation 
has two solutions $k\equiv 1-i-n/2,1-i-n/2+N/2$
for any combination of $i,n/2 \in \mathbb{Z}$ and only one for any $i,n \in 
\mathbb{Z}$, 
if $N$ is odd. By regarding 
\beqn
\Omega{\cal R} \Theta^{1-i-n/2} = \Theta^{-(1-i)/2} 
\left( \Omega{\cal R} \Theta^{-n/2} \right) \Theta^{(1-i)/2}
\eeqn 
one finds that 
$\Omega{\cal R}\Theta^{1-i-n/2}$ leaves the $\left( 6_i, 6_{i+n} \right)$ strings 
invariant just like $\Omega{\cal R}\Theta^{-n/2}$ the $\left( 6_1, 6_{1+n} \right)$ 
strings, which 
are those that start on the brane in the fixed plane of ${\cal R}$. 
On these $\Omega{\cal R}$ itself acts 
simply like the ordinary $\Omega$ on strings with Neumann boundary conditions 
only, as the additional signs of $\Omega$ cancel the reflection. 
The oscillator traces for all other values of $i$ are identical to 
the $i=1$ case, but 
the contributions arise from different $n,k$ combinations. 
For the oscillator part we then get the formula
\beqn
{\cal M}^{(n,k)}_{\left( v_1,v_2,v_3\right)} \equiv (1-1)\ 
\frac{\vartheta \left[ 1/2 \atop 0 \right]}{\eta^3} 
\prod_{(nv_i,kv_i) \notin \mathbb{Z}^2}{ \left( 
        \frac{2^\delta \vartheta \left[ 1/2 + nv_i \atop kv_i \right]}
        {\vartheta \left[ 1/2 + nv_i \atop 1/2 + kv_i \right]}
        e^{\pi i \langle nv_i\rangle}
        \right)
}
\prod_{(nv_i,kv_i) \in \mathbb{Z}^2}{ \left(  
        \frac{\vartheta \left[ 1/2 \atop 0 \right]}{\eta^3} 
        \right)
} 
\eeqn
with argument $q= -\exp\left(-2\pi t\right)$, 
which enforces a different modular transformation property. 
As for the annulus, there are lattice contributions if $nv_i$ and $kv_i$ 
are both integers. They differ from those of the annulus, as one needs 
to sum over states invariant under $\Omega{\cal R}$, which boils down to 
doubling 
the winding quantum numbers, except for the $\mathbb{Z}_4$, where they 
remain unchanged for the ${\bf A}$ type of lattice, while they are still 
doubled for the ${\bf B}$ type. For the modular transformed M\"obius strip 
we use 
\beqn
\tilde{\cal M}^{(m)}_{\left( v_1,v_2,v_3\right)} & \equiv & (1-1)\
\frac{\vartheta \left[ 1/2 \atop 0 \right]}{\eta^3} 
\prod_{mv_i \notin \mathbb{Z}}{ \left( 
        \frac{\vartheta \left[ 1/2 \atop mv_i \right]}
        {\vartheta \left[ 1/2 \atop 1/2+ mv_i \right]}
        \right)
}
\prod_{mv_i \in \mathbb{Z}}{ \left(  
        \frac{\vartheta \left[ 1/2 \atop 0 \right]}{\eta^3} 
        \right)
} 
\eeqn
%\beqn
%\tilde{\cal M}^{(n,k)}_{\left( v_1,v_2,v_3\right)} & \equiv & (1-1)\ 
%\frac{\vartheta \left[ 0 \atop 1/2 \right]
%        \vartheta \left[ 1/2 \atop 0 \right]}
%        {\eta^3 \vartheta \left[ 0 \atop 0 \right]} 
%\prod_{(nv_i,kv_i) \notin \mathbb{Z}^2}{ \left( 
%        \frac{\vartheta \left[ nv_i/2+kv_i \atop (nv_i+1)/2 \right]
%        \vartheta \left[ (nv_i+1)/2+kv_i \atop nv_i/2 \right]}
%        {\vartheta \left[ (nv_i+1)/2+kv_i \atop (nv_i+1)/2 \right]
%        \vartheta \left[ nv_i/2+kv_i \atop nv_i/2 \right]}
%        \right)
%} \times \non 
%& & 
%\prod_{(nv_i,kv_i) \in \mathbb{Z}^2}{ \left(  
%        \frac{\vartheta \left[ 0 \atop 1/2 \right]
%        \vartheta \left[ 1/2 \atop 0 \right]}
%        {\eta^3 \vartheta \left[ 0 \atop 0 \right]} 
%        \right)
%} 
%\eeqn
with argument $\tilde{q}=-\exp\left( -4\pi l\right)$. 

\section{Definitions and modular transformation formulas}
\label{thetafu}

In this appendix we give basic definitions and fix some notation. 
We frequently employ the Jacobi theta function and Dedekind eta function
\beqn
\vartheta \left[ \alpha \atop \beta \right] (t) & = &  
\sum_{n\in \mathbb{Z}}{q^{(n+\alpha)^2 /2}\ e^{2\pi i (n+\alpha) \beta}}, \non
\eta(t) & = & q^{1/24} \prod_{n=1}^\infty{\left( 1-q^n \right)} ,
\eeqn
setting $q \equiv e^{-2\pi t}$. The argument $\alpha$  is 
defined modulo $\mathbb{Z}$ and in order to directly use the product expansion
\beqn
\frac{\vartheta \left[ \alpha \atop \beta \right]}{\eta} (t) = 
e^{2\pi i \alpha \beta}\ q^{\alpha^2/2 -1/24} 
\prod_{n=1}^\infty{\left( \left( 1+q^{n-1/2+\alpha} e^{2\pi i \beta} \right) 
\left( 1+ q^{n-1/2-\alpha} e^{-2\pi i \beta} \right) \right) } 
\eeqn
one needs to  choose $\alpha \in (-1/2, 1/2]$. 

The modular $S$ transformation is 
\beqn
\vartheta \left[ \alpha \atop \beta \right] \left( t^{-1} \right) & = &  
\sqrt{t}\ e^{2\pi i \alpha \beta} 
\vartheta \left[ -\beta \atop \alpha \right] \left( t \right) , \non
\eta \left( t^{-1} \right) & = & \sqrt{t}\ \eta ( t ).
\eeqn
In order to rewrite the loop channel M\"obius strip amplitude in terms of 
the appropriate tree level fields we use the identity
\beqn
\frac{\vartheta \left[ \alpha +1/2 \atop \beta \right]}
        {\vartheta \left[ \alpha +1/2 \atop \beta +1/2 \right]} (-q) =
e^{-\pi i \alpha}
\frac{\vartheta \left[ (\alpha +1)/2 \atop \alpha/2 + \beta \right]
        \vartheta \left[ \alpha /2 \atop (\alpha +1)/2 + \beta \right]}
        {\vartheta \left[ (\alpha +1)/2 \atop (\alpha+1)/2 + \beta \right]
        \vartheta \left[ \alpha /2 \atop \alpha/2 + \beta \right]} \left( q^2 
\right)
\eeqn
for $-1<\alpha\le 0$. 
For the modular transformation of lattice and momentum and winding sums 
we need the Poisson resummation formula
\beqn
\sum_{n \in \mathbb{Z}}{e^{-\pi n^2/t}} = 
\sqrt{t}\ \sum_{n\in \mathbb{Z}}{e^{-\pi n^2 t}} .
\eeqn

\end{appendix}

\clearpage
\bibliography{articles}
\bibliographystyle{unsrt}

\end{document}